\documentclass{mn2e}
\usepackage{epsfig}

\renewcommand{\d}{{\rm d}}
\newcommand{\beq}{\begin{equation}}
\newcommand{\eeq}{\end{equation}}
\newcommand{\beqa}{\begin{eqnarray}} 
\newcommand{\eeqa}{\end{eqnarray}}
\newcommand{\bea}{\begin{array}} 
\newcommand{\ea}{\end{array}} 
\newcommand{\lag}{\langle}
\newcommand{\rag}{\rangle}
\newcommand{\Om}{\Omega_{\rm m}}
\newcommand{\Ol}{\Omega_{\Lambda}}
\newcommand{\inta}{\int_{-i\infty}^{+i\infty}}
\newcommand{\cD}{{\cal D}}

\title[Weak lensing of SNeIa flux distribution]
{How many SNeIa do we need to detect the effect of weak lensing ?}

\author[Munshi \& Valageas]
{Dipak Munshi$^{1,2}$, Patrick Valageas$^{3}$\\
$^{1}$Institute of Astronomy, Madingley Road,
Cambridge, CB3 OHA, United Kingdom\\
$^{2}$Astrophysics Group, Cavendish Laboratory, Madingley Road, 
Cambridge CB3 OHE, United Kingdom\\
$^{3}$Service de Physique Th\'eorique, 
CEA Saclay, 91191 Gif-sur-Yvette, France \\}

\begin{document}
\maketitle

\begin{abstract}
We show that as many as $4000$ SNeIa may be required to detect the effect
of weak lensing on their flux distribution with a high level of significance. 
However, if the intrinsic SNeIa magnitude dispersion is unknown one needs 
an even higher number of SNeIa (an order of magnitude more) to reach a similar 
level of statistical significance. Moreover, the ability to separate the 
lensing contribution from the intrinsic scatter depends sensitively on the 
amplitude of the latter. Using a Kolmogorov - Smirnov (K-S) test we check how 
the required number of SNeIa changes with level of significance. 
Our model incorporates a completely analytical 
description of weak lensing which has been tested extensively against 
numerical simulations. Thus, future missions such as SNAP may be able
to detect non-Gaussianity at a lower significance level of $10\%$ (through the
K-S test) only if the intrinsic scatter is known from external data (e.g. from 
low redshift observations) whereas ALPACA with $100,000$ SNe will definitely 
detect non-Gaussianity with a very high confidence even if the intrinsic 
magnitude dispersion is not known {\it a priori}.
\end{abstract}

\begin{keywords}
Cosmology: theory -- gravitational lensing -- large-scale structure of Universe
Methods: analytical -- Methods: statistical --Methods: numerical
\end{keywords}

\section{Introduction}

Type Ia supernovae (SNeIa) are powerful probes of the
recent expansion of the universe and they provided the main contribution to the
discovery of the present acceleration of the universe (Riess et al. 1998;
Perlmutter et al. 1999). Indeed, SNeIa are standard candles with a small
luminosity dispersion so that by measuring the flux received on the earth 
one can derive the luminosity distance of the source.
Then, by observing many SNeIa one can measure the redshift-distance relation
which provides constraints on cosmological parameters 
(Goobar \& Perlmutter 1995). However, even SNeIa are not perfect candles
and are affected by various sources of noise such as the magnification 
produced by gravitational lensing, 
related to the fluctuations of the matter distribution along the line of 
sight (Kantowski et al. 1995; Frieman 1997). Flux conservation implies that 
the random magnification shift is zero (Weinberg 1976) but weak 
lensing distortions increase the observed SNeIa magnitude scatter and lead
to an extended high-luminosity tail (Wambsganss et al. 1997; Valageas 2000).
For a flux-limited survey weak lensing also leads to a slight bias
towards larger luminosities close to the threshold (Valageas 2000) but this 
plays no significant role. Then, from the deviation of the magnitude
distribution of 63 high redshift SNeIa (Riess et al. 2004) from a Gaussian,
Wang (2005) claimed that weak lensing effects may have been detected.
Although the distortion agrees at a qualitative level with the expectation
from weak lensing magnification (i.e. there are three very bright SNeIa)
the statistics was too small to draw firm conclusions. In this {\it Letter}
we revisit this issue by investigating how many SNeIa are needed to detect
with high confidence weak lensing effects. In 
sect.~\ref{Detecting_weak_lensing} we describe how weak gravitational lensing
by large scale structures affects the apparent magnitude of SNeIa. Next, 
assuming that the intrinsic magnitude fluctuation (including all sources of 
noise except lensing) is Gaussian with a known variance we discuss a 
Kolmogorov-Smirnov test to assess whether a sample of $N$ supernovae may
be drawn from such a Gaussian. Then, in sect.~\ref{Known_intrinsic_variance} 
we compute for a survey such as the proposed 
SNAP\footnote{http://snap.lbl.gov} experiment 
(Aldering et al. 2004) at which $N_*$ a deviation from this Gaussian is 
detected with a high confidence level. In sect.~\ref{Marginalizing} we
generalize this procedure to the case where the intrinsic magnitude variance
is unknown. We discuss the dependence of our results on the amplitude of the 
intrinsic SNeIa magnitude dispersion in 
sect.~\ref{Dependence_on_intrinsic_variance} and we conclude in 
sect.~\ref{Conclusions}.

\section{Detecting weak lensing}
\label{Detecting_weak_lensing}

If there are no distortions the apparent magnitude $m_{\rm app}$ of a 
supernova at redshift $z$ is related to its absolute magnitude $M_*$ and 
luminosity $L_*$ by:
\beq
m_{\rm app} = M_* + 5 \log\left[\frac{d_L(z)}{10 {\rm pc}}\right] + K(z), 
M_* = -2.5 \log\left(\frac{L_*}{L_0}\right) ,
\label{mapp}
\eeq
where $d_L(z)$ is the luminosity distance, $K(z)$ the ``K-correction'' which
describes the redshift of the flux spectrum with respect to
the observing filter and $L_0$ the zero-point of the magnitude system. 
Therefore, inverting eq.(\ref{mapp})
observers who measure the flux from distant supernovae can derive $d_L(z)$
which provides constraints on cosmological parameters. However, in practice
one needs to take into account the intrinsic dispersion $\sigma_{\rm int}$
of supernovae magnitudes, which is due to the dispersion of SNeIa luminosities
themselves as well as to measurement noises and absorption along the line of
sight. This implies a statistical analysis to extract the mean distance
modulus $m_{app}-M_*$. Another distortion is due to weak lensing which
can magnify the luminosity of distant supernovae. Therefore, the apparent
magnitude shows a fluctuation $\delta m$ around its average 
$\lag m_{\rm app}\rag$:
\beq
\delta m = \delta m_{\rm{int}} + \delta m_{\rm{lens}} , \;\;\;
\lag\delta m\rag=0 ,
\label{mappdist}
\eeq
where we separate the intrinsic fluctuation $\delta m_{\rm{int}}$ and the 
gravitational lensing distortion $\delta m_{\rm{lens}}$. Note that since the 
mean weak lensing magnification is unity (as weak lensing only modifies the 
trajectory of light rays and does not change their energy) gravitational
lensing does not bias the average apparent magnitude which allows one to 
derive $d_L$ and to put constraints on cosmology. In this work, we focus on 
the fluctuating part $\delta m$ and we investigate how many SNeIa are 
required to detect weak lensing through the statistics of the fluctuations 
$\delta m$, which depend on $\delta m_{\rm{lens}}$. In the weak lensing 
regime which is appropriate for lensing by large scale structures that we 
consider here the magnification $\mu$ is related to the usual weak 
lensing convergence $\kappa$ by:
\beq
\left. \frac{L_{\rm{obs}}}{L_{\rm{true}}} \right|_{\rm{lens}} = \mu 
\simeq 1+2\kappa ,
\label{mu}
\eeq
which can be written in terms of the density contrast $\delta$ along the
line of sight as:
\beq
\kappa = \frac{3\Om}{2} \frac{H_0^2}{c^2} \int_0^{\chi_s} \d\chi 
\frac{\cD(\chi)\cD(\chi_s-\chi)}{\cD(\chi_s)} (1+z) \delta(z) .
\label{kappa}
\eeq
Here $H_0$ is the Hubble constant, $\chi$ is the radial distance along the
line of sight and $\cD$ the angular diameter distance. From the definition
of magnitudes in (\ref{mapp}) we obtain for the apparent magnitude fluctuation:
\beq
\delta m = \delta m_{\rm{int}} + \delta m_{\rm{lens}} 
= \delta m_{\rm{int}} - \frac{5\kappa}{\ln 10} .
\label{dmapp}
\eeq
We shall assume in the following that $\delta m_{\rm{int}}$ is Gaussian with
variance $\sigma_{\rm{int}}$. Then, as in Valageas et al. (2005) the 
probability distribution of $\delta m$ can be written in terms of 
its generating function $\varphi_{\delta m}$ as:
\beq
P(\delta m) = \inta \frac{\d y}{2\pi i \lag\delta m^2\rag}
e^{[\delta m y - \varphi_{\delta m}(y)]/\lag\delta m^2\rag}
\label{Pdeltam}
\eeq
with:
\beq
\varphi_{\delta m}(y) = \frac{1+\rho}{\rho} \varphi_{\delta m_{\rm{lens}}}
\left( \frac{\rho}{1+\rho} y\right) - \frac{1}{1+\rho} \frac{y^2}{2} .
\label{phiy}
\eeq
Here we introduced the generating function $\varphi_{\delta m_{\rm{lens}}}$ 
of the lensing fluctuation and we defined the ratio $\rho$ of intrinsic
and lensing variances by:
\beq
\rho = \frac{\lag\delta m_{\rm{lens}}^2\rag}{\sigma_{\rm{int}}^2} 
= \left(\frac{5}{\ln 10}\right)^2 \frac{\lag\kappa^2\rag}{\sigma_{\rm{int}}^2}.
\label{rho}
\eeq
Finally, we use the model described in Barber et al. (2004) or 
Munshi et al. (2004) to obtain the generating functions of the convergence
$\kappa$ whence of the lensing magnitude fluctuation $\delta m_{\rm{lens}}$,
taking into account the redshift distribution of the sources (here SNeIa) as
in Valageas et al. (2005).

Then, from the observed distribution of apparent
magnitudes, whence of $\delta m$, one can recover the statistics
of $\kappa$. In this fashion, from the tails of the observed magnitude
distribution Wang (2005) claimed that weak lensing effects may have been
detected. However, the sample was too small (67 high redshift SNeIa) to draw
definite conclusions. Here we reconsider this question by computing how many
SNeIa are needed to get a clear detection of weak lensing from SNeIa.

To this order, we use a Kolmogorov-Smirnov (K-S) test (Press et al. 1986, Kendall \& Stuart 1969)  as follows. From a sample
of $N$ supernovae the observer can compare their magnitude distribution
with a Gaussian $P_G$ of variance $\sigma_G$ through the K-S 
distance $d$ defined by:
\beq
d= \max_{\delta m} \left| S_N(\delta m) - P_G(<\delta m) \right|.
\label{dKS}
\eeq
Here $P_G(<\delta m)$ is the trial cumulative Gaussian whereas 
$S_N(\delta m)$ is the discrete cumulative distribution obtained from the
data. Thus, $S_N(\delta m)$ is merely the fraction of observed SNeIa
with a magnitude fluctuation smaller than $\delta m$. As is well known, the 
interest of the K-S estimator $d$ is that its distribution is universal when 
the data is compared with its parent distribution (null hypothesis), whatever 
it is. Thus, the cumulative distribution of $d$ writes in this case:
\beq
P(>d) = Q_{KS}(\sqrt{N} d) \;\; \mbox{with} \;\; 
Q_{KS}(\lambda)= - 2 \sum_{j=1}^{\infty} (-)^j e^{-2j^2\lambda^2}
\label{QKS}
\eeq
As expected, eq.(\ref{QKS}) shows that the distance $d$ between the discrete
cumulative distribution $S_N$ and its continuous parent distribution scales
as $1/\sqrt{N}$.
Then, by computing the distance $d$ of the sample with respect to a trial
Gaussian from eq.(\ref{dKS}), one can obtain from eq.(\ref{QKS}) the 
probability $P(>d)$ that a distance of this size or larger would be observed
if the trial Gaussian is the true parent magnitude distribution. Therefore, if
this significance level $P(>d)$ is smaller than a threshold $P_- \ll 1$ one
can conclude with good confidence that this trial Gaussian is not the true 
parent distribution (disproof of the null hypothesis). In our case, this means
that weak lensing has been detected since we assume that this is the only 
source of distortion from the Gaussian of intrinsic variance 
$\sigma_G$ with $\sigma_G=\sigma_{\rm int}$.

Thus, to find out for which $N$ such a disproof of the Gaussian can be obtained
with a high significance level we first choose a threshold $P_-\ll 1$ (for
instance $P_-=5\%$) and compute from eq.(\ref{QKS}) the scaled variable 
$\lambda_-$ such that $Q_{KS}(\lambda_-)=P_-$ (for $P_-=5\%$ this yields 
$\lambda_-=1.34$). Then, we draw a large number $N_{\rm{sim}}$ 
of samples of $N$ supernovae magnitude fluctuations $\delta m_i$ ($i=1,..,N$), 
for some value of $N$. As explained above, the distribution of these 
magnitudes is obtained from eq.(\ref{Pdeltam}). Next, we compute for each 
sample $k$ ($k=1,..,N_{\rm{sim}}$) the distance $d_k$ to the Gaussian
of variance $\sigma_G=\sigma_{\rm int}$, using eq.(\ref{dKS}). From this 
set $\{d_k\}$
we obtain the probability $P(>\lambda_-)$ to measure a distance $d$ larger than
our threshold $d_-=\lambda_-/\sqrt{N}$. This is simply the fraction of 
realizations among our $N_{\rm{sim}}$ simulations with $d_k>d_-$. Then, we can
repeat the same procedure for various $N$ which provides the curve 
$P(>\lambda_-;N)$ as a function of $N$ (at fixed $\lambda_-$). Since the parent
distribution (\ref{Pdeltam}) is different from the Gaussian of variance
$\sigma_{\rm int}$ this probability $P(>\lambda_-;N)$ increases with $N$ and 
goes to unity at large $N$: for sufficiently large $N$ we are sure to detect
the difference between both PDF. Finally, we select a second threshold 
$P_+ \simeq 1$ (for instance $P_+=95\%$) and we find above which $N_*$ the
probability $P(>\lambda_-;N)$ becomes larger than $P_+$. This value of $N_*$
is the number of supernovae needed to detect with a high probability ($P_+$)
a weak lensing signature (defined as a distance $d$ from the Gaussian which
is more rare than $P_-$).

\section{Numerical results}
\label{Numerical_results}

We assume a concordance $\Lambda$CDM cosmology with $\Om=0.3$, $\Ol=0.7$, 
$\sigma_8=0.88$ and $H_0=70$km/s/Mpc. We also adopt a redshift distribution
of SNeIa as expected for the SNAP mission (Table~1 of Kim et al. 2004)
which plans to observe about 6000 supernovae, of which 2000 may be used for
cosmological purposes between redshifts of $0.1$ and $1.7$ 
(Aldering et al. 2004). We also use throughout an intrinsic magnitude 
dispersion $\sigma_{\rm int}=0.1$ mag.

\begin{figure}
\begin{center}
\epsfxsize=7 cm \epsfysize=5.3 cm {\epsfbox{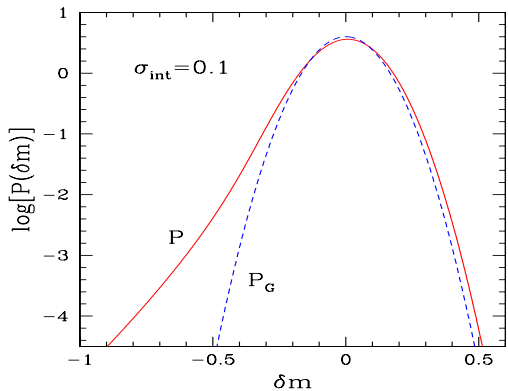}} 
\end{center}
\caption{The probability distribution $P(\delta m)$ (solid line) of the SNeIa 
magnitude fluctuation from the mean, as given by eq.(\ref{Pdeltam}). 
The dashed line shows the Gaussian $P_G$ of variance 
$\sigma_G=\sigma_{\rm int}=0.1$ mag
which corresponds to neglecting weak-lensing effects.}
\label{fig:pdf}
\end{figure}

We show in Fig.~\ref{fig:pdf} the probability distribution $P(\delta m)$ of
the SNeIa magnitude fluctuation $\delta m$ (solid line) from 
eq.(\ref{Pdeltam}). We also plot for comparison the Gaussian $P_G$ of 
variance $\sigma_{\rm int}=0.1$. Thus, we see that weak-lensing effects 
increase the dispersion $\lag(\delta m)^2\rag$ and distort the shape of the
distribution with an extended bright tail.

\subsection{Known intrinsic variance}
\label{Known_intrinsic_variance}

\begin{figure}
\begin{center}
\epsfxsize=6. cm \epsfysize=6.5 cm {\epsfbox[155 12 400 276]{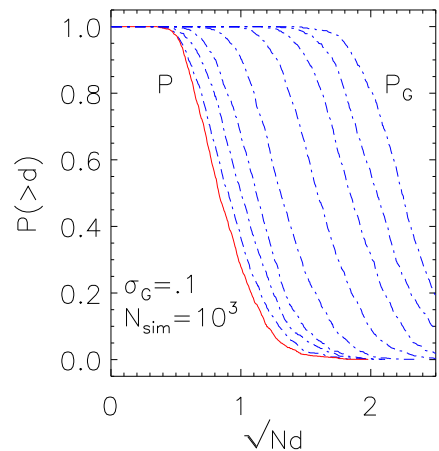}} 
\end{center}
\caption{The cumulative probability distribution $P(>\lambda)$ to measure
a distance larger than $d=\lambda/\sqrt{N}$ from the Gaussian of variance 
$\sigma_G=\sigma_{\rm int}=0.1$ mag. The dot-dashed curves correspond to 
$N=100,~ 250,~500,~1000,~2000,~3000,~4000,~5000$ from left to right.
The left solid curve shows for reference the cumulative probability 
distribution of the distance from the parent distribution (\ref{Pdeltam}).
It is equal to $Q_{KS}$ in eq.(\ref{QKS}).}
\label{fig:allNcpdf}
\end{figure}

We first apply in this section the K-S test as described above in 
sect.~\ref{Detecting_weak_lensing}. Thus, we show in Fig.~\ref{fig:allNcpdf}
the cumulative probability distribution $P(>\lambda)$ to measure
a distance larger than $d=\lambda/\sqrt{N}$ from the Gaussian of variance 
$\sigma_G=\sigma_{\rm int}$. These PDF are obtained for each $N$ from the 
distribution of distances $\{d_k\}$ ($k=1,..,N_{\rm sim}$) associated with our 
$N_{\rm sim}$ realizations of $N$ supernovae. As $N$ increases the 
cumulative probability $P(>\lambda)$ develops a plateau which extends to
larger values of $\lambda$ as it is easier to detect the deviation of the
parent distribution (\ref{Pdeltam}) from the trial Gaussian of variance
$\sigma_{\rm int}$. Therefore, the probability $P(>\lambda_-)$ grows with $N$.

\begin{figure}
\begin{center}
\epsfxsize=6. cm \epsfysize=6.5 cm {\epsfbox[155 12 400 276]{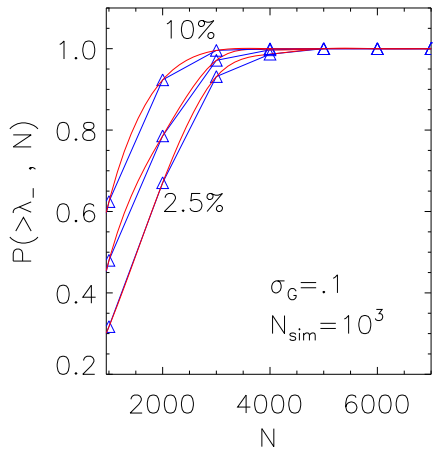}} 
\end{center}
\caption{The curves $P(>\lambda_-;N)$ for three different thresholds 
$P_-=10\%,5\%$ and $2.5\%$ (i.e. $\lambda_-=1.20, 1.34$ and $1.46$) from top 
downto bottom, as a function of $N$. At large $N$ one detects almost surely 
($P(>\lambda_-;N)\simeq 1$) a large deviation from the trial Gaussian 
(so large that it would have occurred with probability $P_-$ if the latter 
Gaussian had been the true parent distribution).}
\label{fig:signif}
\end{figure}

Thus, we show in Fig.~\ref{fig:signif} the curves $P(>\lambda_-;N)$ as a 
function of $N$, for three different thresholds $P_-=10\%, 5\%$ and $2.5\%$
(corresponding to $\lambda_-=1.20, 1.34$ and $1.46$) from top downto bottom. We 
can check that for low $N$ the statistics is too small to obtain a clear 
detection of weak lensing and as $N$ increases the probability to measure the
deviation from the Gaussian due to weak lensing effects grows to reach unity
at $N\rightarrow \infty$. Thus, we find that for significance levels
$\{P_-=10\%,P_+=90\%\}$, $N_*=2000$ supernovae are sufficient to detect
weak lensing. Higher levels $\{P_-=5\%,P_+=95\%\}$ and 
$\{P_-=2.5\%,P_+=97.5\%\}$ require $N_*=3000$ and $N_*=4000$ supernovae.

\subsection{Marginalizing over observed variance}
\label{Marginalizing}

\begin{figure}
\begin{center}
\epsfxsize=5.7 cm \epsfysize=6.3 cm {\epsfbox[155 12 400 276]{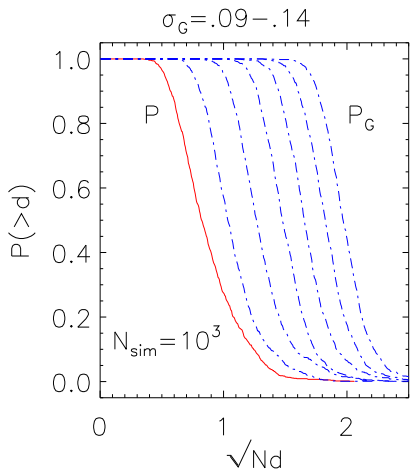}} 
\end{center}
\caption{Same as Fig.~\ref{fig:allNcpdf} but with marginalization over 
the observed variance. 
The dot-dashed curves show for various $N$ the cumulative probability 
distribution $P(>\lambda_{\rm min})$ to measure a distance larger than
$d_{\rm min}=\lambda_{\rm min}/\sqrt{N}$ from the closest Gaussian among
Gaussians of any variance (it is sufficient to span the range 
$0.09<\sigma_G<0.14$).
They correspond to $N=10000,20000,30000,40000,50000,60000$ from left to right. 
The left solid curve shows for reference the distance from the parent 
distribution (\ref{Pdeltam}) and obeys eq.(\ref{QKS}). For each of these 
studies statistics are constructed from $1000$ simulations.
The intrinsic variance is again $\sigma_{int}=0.1$ mag.}
\label{fig:allNcpdf_marg}
\end{figure}

The procedure used in the previous section assumes that the intrinsic
variance $\sigma_{\rm int}$ is exactly known so that any deviation from
the Gaussian of variance $\sigma_{\rm int}$ is interpreted as a detection
of weak lensing. However, in practice the variance $\sigma_{\rm int}$ is only
known up to some finite accuracy. Moreover, high redshift SNeIa may exhibit
a somewhat different variance (because of the evolution of SNeIa metallicities,
absorption by dust along the line of sight, etc.). Therefore, in this section
we marginalize over the variance of the observed sample (keeping
$\sigma_{\rm int}=0.1$ mag for the true parent distribution), which implies 
that detection of weak lensing only depends on non-Gaussianities. 
Thus, for each realization $k$ of
$N$ supernovae we compute all distances $d_{k;p}$ of this data set from
an ensemble of trial Gaussians $P_{G;p}$ of different variances $\sigma_{G;p}$.
From these $d_{k;p}$ we obtain the minimum distance 
$d_{{\rm min};k}=\min_p \{d_{k;p}\}$. Thus $d_{{\rm min};k}$ is the minimum
distance between this realization and any Gaussian. In practice we use a grid
of variances $\sigma_{G;p}$ which spans the range $[0.09,0.14]$ with a step of
$0.002$. Obviously the distance $d_{k;p}$ increases at very small or very
large variance $\sigma_G$ and it is minimum for 
$\sigma_G^2 \simeq \sigma_{\rm int}^2 + \sigma_{\rm lens}^2$. For the 
cosmology and the redshift distribution that we use in this work 
we find that the minimum distance corresponds to $\sigma_G \simeq 0.125$.
Then, from the distribution of minimum distances $d_{{\rm min};k}$ provided
by our $N_{\rm sim}$ realizations of $N$ supernovae magnitudes 
we obtain the cumulative probability distribution $P(>d_{\rm min})$ to observe
a distance larger than $d_{\rm min}$ from the closest possible Gaussian 
distribution. We show in Fig.~\ref{fig:allNcpdf_marg} this cumulative 
probability distribution. Of course, we can check that for a given $N$
the typical distance $d_{\rm min}$ is smaller than the distance $d$ to the
fixed Gaussian of variance $\sigma_G=\sigma_{\rm int}$ used 
in Fig.~\ref{fig:allNcpdf}.
Therefore, a larger number of SNeIa is needed to detect weak lensing.

\begin{figure}
\begin{center}
\epsfxsize=6. cm \epsfysize=6.5 cm {\epsfbox[155 12 400 276]{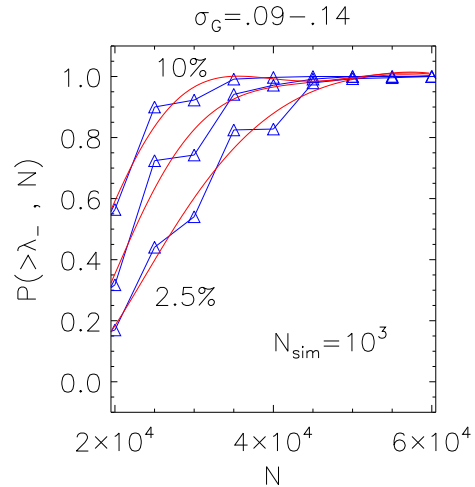}} 
\end{center}
\caption{The curves $P(>\lambda_-;N)$ for three different thresholds 
$P_-=10\%,5\%$ and $2.5\%$ from top downto bottom as in 
Fig.~\ref{fig:signif} but using the distance to the closest Gaussian displayed
in Fig.~\ref{fig:allNcpdf_marg}. 
A range of Gaussian PDFs were used (see text for more details). 
Triangles are actual estimates from our simulations whereas the solid lines 
are fitting functions.}
\label{fig:signif_marg}
\end{figure}

Applying the same procedure as in sect.~\ref{Known_intrinsic_variance} we
can now display in Fig.~\ref{fig:signif_marg} the curves $P(>\lambda_-;N)$
as a function of $N$ obtained from the minimum distance distributions shown
in Fig.~\ref{fig:allNcpdf_marg}. We see that an order of magnitude more SNe 
are needed if the intrinsic variance is not known in advance (e.g. from 
low redshift studies). About $50,000$ SNe are required to detect weak lensing
effects in SNeIa studies through the K-S test with a high level of confidence 
for the redshift distribution that we have considered here. In particular,
we now find that the significance levels
$\{P_-=10\%,P_+=90\%\}$, $\{5\%,95\%\}$ and $\{2.5\%,97.5\%\}$
require $N_*=30,000, 40,000$ and $45,000$ supernovae.
Note that the SNAP mission actually plans to observe $\sim 10,000$
SNeIa from which $\sim 4000$ should be well-characterized (Albert et al. 2005).
Therefore, it will be able to detect weak lensing only if the intrinsic 
dispersion of SNeIa magnitudes is known.
On the other hand, the JEDI\footnote{http://jedi.nhn.ou.edu} experiment 
proposes to observe over $14,000$ SNeIa
with well sampled light curve and good quality spectra (Crotts et al. 2005)
over $0<z<1.7$ whereas the 
ALPACA\footnote{http://www.astro.ubc.ca/LMT/alpaca/index.html} experiment 
plans to observe $\ga 100,000$
supernovae in the range $0.2<z<1$ (Corasaniti et al. 2005).

\begin{figure}
\begin{center}
\epsfxsize=6. cm \epsfysize=6.5 cm {\epsfbox[155 12 400 276]{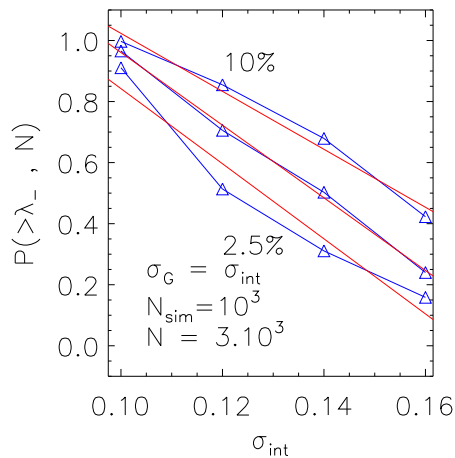}} 
\end{center}
\caption{The curves $P(>\lambda_-;N)$ for three different thresholds 
$P_-=10\%,5\%$ and $2.5\%$ from top downto bottom. All simulations were
performed with $3000$ SNe. A set of $1000$ simulations were performed to 
reduce the scatter. Solid lines represent a linear fit in the range of
intrinsic variances considered in this study 
$\sigma_{int} = .1,.12,.14,.16$. The variance of the trial Gaussian 
distribution is taken equal to the intrinsic variance
$\sigma_G=\sigma_{\rm int}$ for this study (case of known intrinsic variance).
}
\label{fig:signif_change}
\end{figure}

\subsection{Dependence on intrinsic variance}
\label{Dependence_on_intrinsic_variance}

In these studies we have assumed that the intrinsic variance (which can be 
unknown) is $\sigma_{\rm int}=0.1$ mag. However these results are quite
sensitive to $\sigma_{\rm int}$. In Fig.~\ref{fig:signif_change} we plot
for the case of $N=3000$ SNeIa the cumulative probability $P(>\lambda_-;N)$ 
as a function of $\sigma_{\rm int}$. We consider the three thresholds used
in Figs.~\ref{fig:signif}, \ref{fig:signif_marg}, and we use 
$1000$ simulations. As in sect.~\ref{Known_intrinsic_variance} we consider
the case where the intrinsic variance is known so that the observed SNeIa
sample is compared with the trial Gaussian of variance 
$\sigma_G=\sigma_{\rm int}$. Of course, for low $\sigma_{\rm int}$ it is easy
to detect weak lensing (the probability $P(>\lambda_-;N)$ goes to unity)
since the amplitude of weak lensing effects becomes larger than the intrinsic
dispersion of SNeIa magnitudes whereas for high $\sigma_{\rm int}$
weak lensing distortions become relatively negligible ($P(>\lambda_-;N)$ goes 
to zero). We can see that this probability $P(>\lambda_-;N)$
is quite sensitive to $\sigma_{\rm int}$ as it exhibits a fast decrease for
larger $\sigma_{\rm int}$. This implies that the number of SNeIa required to 
detect weak lensing signatures through the Kolmogorov-Smirnov test grows
quickly with the intrinsic SNeIa magnitude variance.
In particular, for $\sigma_{\rm int}=0.16$ mag we find that $7000$, $10000$
and $20000$ SNeIa are required in order to achieve the confidence levels
$\{10\%,90\%\}$, $\{5\%,95\%\}$ and $\{2.5\%,97.5\%\}$ (in the case of known
intrinsic variance as in sect.~\ref{Known_intrinsic_variance}).

\section{Conclusions And Outlook}
\label{Conclusions}

In this {\it Letter} we have addressed the issue of determining the number of
observed SNeIa beyond which weak lensing effects can be detected with 
a high confidence. For the concordance $\Lambda$CDM cosmology, using a model 
of the large-scale matter distribution which has
been checked against numerical simulations, we found that $4000$ SNeIa
are necessary to distinguish a weak lensing signature with a significance 
level of $2.5\%$ through a Kolmogorov-Smirnov test. To reach a significance 
level of $10\%$ we only need $2000$ SNeIa. This procedure compares the 
magnitude distribution
of the observed SNeIa with a Gaussian of fixed variance, assuming that the
latter describes all sources of noise except for weak lensing magnification.
If we consider the variance to be a free parameter (e.g. the intrinsic 
SNeIa magnitude dispersion or the instrumental noise are not accurately
known beforehand) we find that $45,000$ supernovae are required to detect
with a high confidence (at a $2.5\%$ level) non-Gaussian signatures. 
Therefore, future experiments
such as those planned within the Joint Dark Energy Mission will exhibit
clear weak lensing signatures if the intrinsic magnitude dispersion of SNeIa
is well known. However to be more confident without any
{\it a priori} knowledge of $\sigma_{\rm int}$ we will have to wait for 
surveys such as ALPACA. 
Of course, the possibility of detecting weak lensing effects on SNeIa 
magnitude distributions also implies that such gravitational lensing
effects should be taken into account or used as a complementary tool 
to constrain cosmology (e.g., Dodelson \& Vallinotto 2005).

\section*{acknowledgments}
DM acknowledges the support from PPARC of grant
RG28936.

\end{document}